\documentclass[longauth]{aa}  

\usepackage{graphicx}
\usepackage{txfonts}
\usepackage{lipsum}
\usepackage{subcaption}         % necessary for continued figures, example in section 3
\usepackage{lscape}             % to rotate a single page table, example in appendix.
\usepackage{placeins}           % useful with \FloatBarrier, to keep 
\begin{document}

%%%%%%%%%%%%%%%%%%%%%%%%%%%%%%%%%%%%%%%%
% if you use custom commands in your title,
% ensure to check your title when submitting!
%%%%%%%%%%%%%%%%%%%%%%%%%%%%%%%%%%%%%%%%

   \title{Probing the magnetic field of a coronal mass ejection with PSR~J1022+1001}

   % \subtitle{A comparison between line-of-sight magnetic field measurements and semi-empirical CME modeling}

%%%%%%%%%%%%%%%%%%%%%%%%%%%%%%%%%%%%%%%%
% Please separate each author with the \and command
%
% Please do not include ORCIDs next to author names.
% Only ORCIDs authenticated by individual authors in EDPS
% editorial system will be taken into account.
% ORCIDs included here will be removed.
%%%%%%%%%%%%%%%%%%%%%%%%%%%%%%%%%%%%%%%%

\author{
    El Mehdi Zahraoui\inst{\ref{aff:auckland_university}}\thanks{elmehdi.zahraoui@autuni.ac.nz}
    \and Hannah T. Rüdisser\inst{\ref{aff:aswo},\ref{aff:graz_university}}\thanks{hannah@ruedisser.at}
    \and Golam M. Shaifullah\inst{\ref{aff:milano_university},\ref{aff:INAF},\ref{aff:INFN}}\thanks{golam.shaifullah@unimib.it}
    \and Caterina Tiburzi\inst{\ref{aff:INAF}}
    \and Jean-Mathias Grießmeier\inst{\ref{aff:LPC2E},\ref{aff:ORN}}
    \and Ute V. Amerstorfer\inst{\ref{aff:aswo}}
    \and Christian Möstl\inst{\ref{aff:aswo}} % christian.moestl@geosphere.at
    \and Mateja Dumbović\inst{\ref{aff:zagreb_university}}
    \and Emma E. Davies\inst{\ref{aff:aswo}} % emma.davies@geosphere.at
    \and Pietro Zucca\inst{\ref{aff:ASTRON}}
    \and Joris P. W. Verbiest\inst{\ref{aff:florida_university},\ref{aff:max-planck_radio},\ref{aff:bielefeld_university}}
    \and Andreas J. Weiss\inst{\ref{aff:maryland_university},\ref{aff:NASA_Goddard}}
    \and Louis Bondonneau\inst{\ref{aff:LPC2E}}
    \and Baptiste Cecconi\inst{\ref{aff:LIRA},\ref{aff:ORN}}
    \and Benedetta Ciardi\inst{\ref{aff:max-planck_astro}}
    \and Christian Vocks\inst{\ref{aff:AIP}}
    \and Gilles Theureau\inst{\ref{aff:LPC2E},\ref{aff:ORN}}
    \and Julien Girard\inst{\ref{aff:LIRA}}
    \and Oleksandr Konovalenko\inst{\ref{aff:NAS_Ukraine}}
    \and Vyacheslav Zakharenko\inst{\ref{aff:NAS_Ukraine}}
    \and Oleg Ulyanov\inst{\ref{aff:NAS_Ukraine}}
    \and Peter Tokarsky\inst{\ref{aff:NAS_Ukraine}}
    \and Stéphane Corbel\inst{\ref{aff:paris-cite}}
    \and Philippe Zarka\inst{\ref{aff:LIRA},\ref{aff:ORN}}
    \and Cyril Tasse\inst{\ref{aff:LUX},\ref{aff:RATT},\ref{aff:ORN}}
    \and Ralf-Jürgen Dettmar\inst{\ref{aff:bochum}}
    \and Ihor P. Kravtsov\inst{\ref{aff:NAS_Ukraine}}
}

\institute{
         {Institute for Radio Astronomy \& Space Research, Auckland University of Technology, Private Bag 92006, Auckland 1142, New Zealand. \label{aff:auckland_university}}
    \and {Austrian Space Weather Office, GeoSphere Austria, Reininghausstra{\ss}e 3, 8020 Graz, Austria. \label{aff:aswo}}
    \and {Institute of Physics, University of Graz, Universit\"atsplatz 5, 8010 Graz, Austria. \label{aff:graz_university}}
    \and {Dipartimento di Fisica ``G. Occhialini'', Universit\'a degli Studi di Milano-Bicocca, Piazza della Scienza 3, 20126 Milano, Italy. \label{aff:milano_university}}
    \and {INAF - Osservatorio Astronomico di Cagliari, via della Scienza 5, 09047 Selargius (CA), Italy. \label{aff:INAF}}
    \and {INFN, Sezione di Milano-Bicocca, Piazza della Scienza 3, 20126 Milano, Italy. \label{aff:INFN}}
    \and {LPC2E, OSUC, University Orleans, CNRS, CNES, Observatoire de Paris, 45071 Orleans, France. \label{aff:LPC2E}}
    \and {Observatoire Radioastronomique de Nan\c{c}ay (ORN), Observatoire de Paris-PSL, CNRS, Universit\'e d'Orl\'eans, Nan\c{c}ay, France. \label{aff:ORN}}
    \and {University of Zagreb, Faculty of Geodesy, Hvar Observatory, Kaciceva 26, Zagreb, Croatia. \label{aff:zagreb_university}}
    \and {ASTRON, Netherlands Institute for Radio Astronomy, Oude Hoogeveensedijk 4, 7991 PD, Dwingeloo, The Netherlands. \label{aff:ASTRON}}
    \and {Florida Space Institute, University of Central Florida, 12354 Research Parkway, Partnership 1 Building, Suite 214, Orlando, FL 32826-0650, USA. \label{aff:florida_university}}
    \and {Max-Planck-Institut f\"ur Radioastronomie, Auf dem H\"ugel 69, 53121 Bonn, Germany. \label{aff:max-planck_radio}}
    \and {Fakult\"at f\"ur Physik, Universit\"at Bielefeld, Postfach 100131, 33501 Bielefeld, Germany. \label{aff:bielefeld_university}}
    \and {Goddard Planetary Heliophysics Institute, University of Maryland, Baltimore County, Baltimore, MD 21250, USA. \label{aff:maryland_university}}
    \and {Heliophysics Science Division, NASA Goddard Space Flight Center, Greenbelt, MD, USA. \label{aff:NASA_Goddard}}
    \and {LIRA, Observatoire de Paris, Universit\'e PSL, Sorbonne Universit\'e, Universit\'e Paris Cit\'e, CY Cergy Paris Universit\'e, CNRS, Meudon, France. \label{aff:LIRA}}
    \and {Max-Planck-Institut for Astrophysics, Karl-Schwarzschild-Stra{\ss}e 1, 85748 Garching, Germany. \label{aff:max-planck_astro}}
    \and {Leibniz Institute for Astrophysics Potsdam (AIP), An der Sternwarte 16, 14482 Potsdam, Germany. \label{aff:AIP}}
    \and {Institute of Radio Astronomy of the National Academy of Sciences of Ukraine, Mystetstv St. 4, Kharkiv, Ukraine 61002. \label{aff:NAS_Ukraine}}
    \and {Universit\'e Paris Cit\'e and Universit\'e Paris-Saclay, CEA, CNRS, AIM, 91190 Gif-sur-Yvette, France. \label{aff:paris-cite}}
    \and {LUX, Observatoire de Paris, Universit\'e PSL, CNRS, 792190 Meudon, France. \label{aff:LUX}}
    \and {Centre for Radio Astronomy Techniques and Technologies (RATT), Department of Physics and Electronics, Rhodes University, Makhanda, 6140, South Africa. \label{aff:RATT}}
    \and {Ruhr-Universit\"at Bochum, Astronomisches Institut, Universit\"atsstr. 150, D-44801 Bochum, Germany. \label{aff:bochum}}
}
   \authorrunning{Zahraoui et al.}
   \titlerunning{Probing CMEs with pulsars and 3DCORE}
   \date{Received \today}

  \abstract
  {
   {
    Coronal mass ejections (CMEs) are major drivers of space weather, yet their internal magnetic field structure remains difficult to constrain from remote sensing alone. Faraday rotation provides sensitivity to CME magnetic fields, but quantitative interpretation is often limited by the lack of independent line-of-sight (LoS) density information.
    }
   {
   We investigate whether low-frequency pulsar observations can provide LoS magnetic field estimates and whether these are consistent with synthetic LoS signatures extracted from a three-dimensional CME reconstruction constrained by Solar Orbiter data.
   } 
   {
   We analyze a CME occultation of the LoS to PSR~J1022+1001 on 20 August 2021 at a projected heliocentric distance of $24.6\,R_\odot$, observed simultaneously with LOFAR and NenuFAR. From LOFAR, we derive time-resolved dispersion measure (DM) and rotation measure (RM) and isolate the CME contributions using background estimates for interstellar, solar wind and ionospheric components. We then infer the density-weighted LoS-averaged magnetic field component $\langle B_{\parallel}\rangle_{\rm PSR}$ from the ratio $\Delta{\rm RM}/\Delta{\rm DM}$. In parallel, we reconstruct the CME using a semi-empirical 3DCORE model fitted to Solar Orbiter in situ magnetic field observations at 0.65~au. We sample the modeled magnetic field along the pulsar LoS using fixed spatial sampling points and compute synthetic LoS-averaged signatures $\langle B_{\parallel}\rangle_{\rm 3D}$ for different flux rope configurations.
   }
   {
   The derived $\langle B_{\parallel}\rangle_{\rm PSR}$ increases from approximately -9~nT to a peak near 63~nT during the observed interval. Comparison with synthetic signatures shows that the polarity and temporal evolution of the LoS signal are strongly dependent on the flux rope configuration and only a South-West-North (SWN) configuration (confirmed by Solar Orbiter in situ data) reproduces the observed sign and overall evolution, whereas alternative configurations are incompatible. The modeled amplitudes, however, are systematically larger than the pulsar-derived values by roughly a factor of five.
   }
   {
   We show that simultaneous low-frequency pulsar DM and RM measurements can provide LoS magnetic field estimates for a CME and can be used to test CME magnetic structure against data-constrained three-dimensional reconstructions. 
   }
   }

               
\maketitle
    \nolinenumbers

\section{Introduction}
Coronal mass ejections \citep[CMEs,][]{howard_secchi} are large-scale eruptions of plasma and magnetic fields from the solar corona and the primary drivers of severe space weather \citep{kilpua_2017_geoeffective}. While white-light observational techniques have been routinely employed to identify and track CMEs, these observations primarily constrain electron density through Thomson scattering and do not provide direct information about the internal magnetic field structure \citep{Tousey_1973, wood_2023_sensingcmemagnetic, gopalswamy_multiview_2024}. Yet it is the magnetic field, particularly the orientation and strength of the north–south $B_Z$ component relative to the Earth’s magnetic field, that largely determines CME geoeffectiveness and remains the central challenge for space weather forecasting \citep{gonzalez_1987, kilpua_2019_forecasting, vourlidas_2019_predicting, wood_2023_sensingcmemagnetic}.

Many CMEs and their interplanetary counterparts (ICMEs) are thought to contain coherently twisted magnetic field structures, often described as magnetic flux ropes, which bind and transport the plasma within them \citep[e.g.,][]{bothmer_1998,cane2003interplanetary,zurbuchen2006situ}. The orientation and handedness of these flux ropes \citep[often referred to as the flux rope type, e.g.,][]{bothmer_1998,mulligan_1998} largely determine the profile of the $B_Z$ component of the interplanetary magnetic field near Earth, which controls geomagnetic activity \citep[e.g.,][]{gonzalez_1987}.

Determining CME magnetic field structure remotely is therefore a major challenge in heliophysics and space weather research. \textit{In situ} spacecraft measurements provide direct magnetic field observations, but sample only a single trajectory through an event and typically offer limited warning time, as continuous solar wind monitoring is currently available primarily near the Sun–Earth Lagrange Point 1 (L1). Recent studies emphasize that progress toward reliable CME magnetic field prediction requires either new missions monitoring the solar wind upstream of Earth \citep[e.g.,][]{lugaz_2024_neednearearthmultispacecraft, weiler_2024, lugaz_2025_needsubl1space, palmerio_2025_monitoringsolarwind, davies_2025_farupstream,gopalswamy_multiview_2024,Caspi2023COMPLETE} or new remote-sensing diagnostics capable of constraining magnetic fields while CMEs propagate through the inner heliosphere \citep[e.g.,][]{Temmer_2023, wood_2023_sensingcmemagnetic}.

Faraday rotation of linearly polarized radio signals has long been recognized as one of the most promising techniques for probing coronal and heliospheric magnetic fields. Faraday rotation measurements are sensitive to CME flux rope properties, including magnetic field strength, orientation, size, velocity, and expansion rate \citep[e.g.,][]{jensen_2008_faradayrotationobservations, jensen_2010_faradayrotationresponse}. Modern Faraday rotation studies demonstrate that combining Faraday rotation with independent density information can significantly improve constraints on CME magnetic structure \citep[e.g.,][]{kooi_2022_modernfaradayrotation}. However, single line-of-sight (LoS) observations introduce ambiguities in determining flux rope orientation, handedness, and azimuthal field direction \citep{jensen_2010_faradayrotationresponse}. Modeling and observational studies have shown that multiple LoS can resolve these ambiguities and significantly improve magnetic field reconstruction \citep{liu2008reconstruction,jensen_2010_faradayrotationresponse,kooi_2017_vlameasurementsfaraday}. 

Important progress has been made in combining Faraday rotation with other observational constraints. For example, \cite{wood_inferences_2020} demonstrated combining stereoscopic imaging, Faraday rotation measurements and \textit{in situ} plasma data to infer CME magnetic structure. More recent work has also shown that Faraday rotation observations can also be used to refine global MHD models and improve CME forecasting \citep{mancuso_2025_enhancingmhdmodel}. Despite these advances, most Faraday rotation studies rely on sources for which independent density measurements along the same LoS are unavailable. As a result, density is usually determined using simplified assumptions, empirical models or auxiliary observations, where the choice of approach can significantly affect results \citep[e.g.,][]{kooi_2021_vlameasurementsfaraday, kooi_2022_modernfaradayrotation}.

One promising approach to overcome these limitations is the monitoring of compact background radio sources, such as galactic pulsars. Pulsars are rapidly rotating highly magnetized compact objects that emit coherent beamed emission along their magnetic poles. Visible mainly at radio wavelengths, these beams sweep across Earth as the pulsar rotates, generating `pulses' at radio telescopes. Before reaching an observer, this emission travels across several magnetized and ionized media: the ionized interstellar medium (IISM), the solar wind (SW) and the terrestrial ionosphere, and may undergo several propagation effects such as dispersion and Faraday rotation. 

Dispersion probes the integrated column density of free electrons along the LoS and is expressed as the dispersion measure (DM). Faraday rotation probes the integrated product of electron density and the magnetic field component parallel to the LoS and is quantified through the rotation measure (RM). The combination of these two diagnostics therefore enables the estimation of the average magnetic field along the propagation path.

Because DM and RM probe electron density and magnetic field simultaneously, pulsars differ fundamentally from most traditional Faraday rotation background sources, which typically lack independent density information along the same LoS. As a result, pulsar observations have been widely used to study the IISM \citep{Counselman_1972,Tiburzi_2019,donner2020dispersion}. However, in solar and space weather research, pulsars have traditionally played a more limited role, mainly serving as secondary sources for interplanetary scintillation (IPS) studies, with quasars typically preferred \citep{Armstrong_1978}. The first modern detection of CME Faraday rotation using a pulsar was reported by \cite{howard_2016_measuringmagneticfield}, who demonstrated the potential of combining RM and DM measurements to derive an upper limit on the magnetic field. More recent pulsar timing studies have further shown that CME passages can produce detectable DM signatures, highlighting the potential of pulsars as probes of heliospheric transients \citep{chowdhury_2026_effectscoronalmass}. 

In this work, we report the occultation of the LoS to pulsar J1022+1001 by a CME that erupted on August 20, 2021. During this event, a serendipitous simultaneous observation was obtained with the European low-frequency interferometers LOFAR and NenuFAR. Variations in dispersion and Faraday rotation observed in the pulsar signal are used to infer the CME's internal electron density and magnetic field along the LoS. We compare these pulsar-derived quantities with LoS–averaged values derived from a three-dimensional reconstruction of the CME using the semi-empirical 3D Coronal Rope Ejection (3DCORE) model \citep{Mostl2018, Weiss2021a, Rudisser2024}, constrained by remote-sensing observations from white-light and extreme ultraviolet imagers and \textit{in situ} measurements from Solar Orbiter (SolO). To our knowledge, this is the first study to directly compare simultaneous low-frequency pulsar-derived LoS magnetic signatures with synthetic signatures extracted from a data-constrained three-dimensional CME reconstruction, providing a direct demonstration of how pulsar-based remote sensing can be used to test CME magnetic field structure through comparison with forward models.

In Section \ref{sec:data}, we provide an overview of the available data, followed by a detailed description of the analysis and modeling methods in Section \ref{sec:methods}. Section \ref{sec:results} presents the results, which are then discussed in Section \ref{sec:discussion}. Finally, Section \ref{sec:conclusions} offers our conclusions and discusses future prospects.

%%%%%%%%%%%%%%%%%%%%%%%%%%%%%%%%%%%%%%%%%%%%%%%%%%%%%%%%%%%%%%
\section{Data}\label{sec:data}

In this section, we describe the different kinds of data used in this article: pulsar observations obtained with the interferometers LOFAR and NenuFAR, remote-sensing observations from different sources and \textit{in situ} data from SolO. 

\subsection{Radio-frequency observations of PSR~J1022+1001}

The radio observations of millisecond pulsar J1022+1001, located at $\text{RAJ} = 10^\mathrm{h} 22^\mathrm{m} 57.9996^\mathrm{s}$, $\text{DECJ} = +10^\circ 01' 52.80''$, used in this study were collected with the European low-frequency interferometers LOFAR \citep{Stappers11,van2013lofar} and NenuFAR \citep[New extension in Nan\c cay upgrading LOFAR, see][]{ZarkaNenufarIEEE,ZarkaNenuFARinstrument22}. A detailed description of pulsar J1022+1001 parameters is available in the ATNF Pulsar Catalogue \citep{manchester2005}.

The targeted pulsar is part of regular monitoring programs with the LOFAR core and NenuFAR. In particular, the LOFAR core has observed this source since 2012, with a monthly to fortnightly cadence and an integration time varying from 10 to 40 minutes \citep{donner2020dispersion}. Besides the regular monitoring, PSR~J1022+1001 was the target of several special campaigns, such as the ones targeting low-ecliptic pulsars during their solar approaches. With NenuFAR, it has been observed since 2019 using the dedicated pulsar instrumentation UnDysPuTeD and associated software LUPPI \citep{Bondonneau21}, with a $\sim14$ days cadence and an integration time of one to two hours. During the 2021 solar conjunction, the cadence was strongly increased for both LOFAR and NenuFAR (daily observations when the pulsar was near the solar conjunction with a two hour integration time).

The observations collected during these monitoring campaigns are collected in full-Stokes, coherently dedispersed and folded 
%using the \textsc{dspsr} (REF) software suite 
(the NenuFAR data are also coherently Faraday-derotated) and stored in \textsc{timer} or \textsc{psrfits} format files \citep{Hotan_2004} referred to as \textit{archives}. These archives have a time and frequency resolution of 10~s and 195~kHz, respectively. While the observations collected with the LOFAR core cover a frequency bandwidth of 78.51\,MHz with a central frequency of 149.12\,MHz divided into 400 frequency channels, those collected with NenuFAR span a bandwidth of 75\,MHz centred at 49.12\,MHz and divided into 384 frequency channels.

\subsection{Remote CME observations}

\begin{figure*}
\centering
\includegraphics[width=0.8\textwidth]{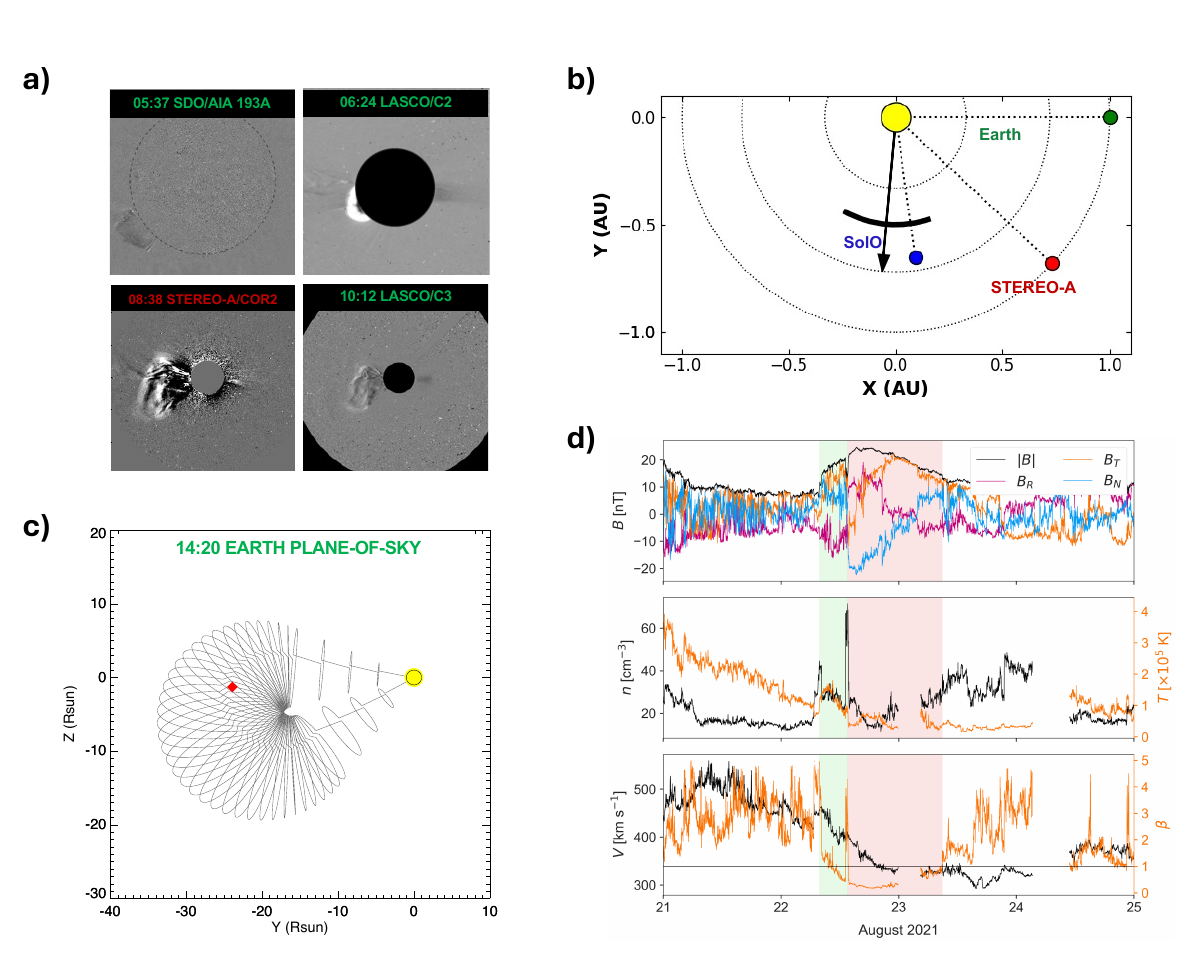}
\caption{Overview of the CME/ICME measurements. a) Early CME measurements as observed in running difference images on the limb in SDO/AIA wavelength 193 \AA{} and later in SOHO/LASCO (along the Sun-Earth line) and STEREO-A coronagraphs. b) Spacecraft configuration at the time of CME detection, as well as the direction and extent of the CME as derived from the GCS reconstruction. c) Mesh of the GCS reconstructed 3D CME geometry as viewed in the Earth plane-of-sky, along with the position of pulsar at the time of pulsar measurement (red dot). d) \textit{In situ} measurements of the ICME detected by SolO in local spacecraft Radial-Tangential-Normal (RTN) coordinates, where the ICME sheath and flux rope are shaded green and red, respectively.}
\label{fig:CME-ICME}
\end{figure*}

We analyze remote-sensing solar observations shown in Figure~\ref{fig:CME-ICME}a) to identify the CME responsible for the signal in radio measurements. Specifically, we use white-light coronagraph data from two vantage points. From the Sun-Earth line, we utilize the C2 and C3 coronagraphs of the Large Angle and Spectrometric Coronagraph Experiment \citep[LASCO;][]{brueckner_lasco} onboard the Solar and Heliospheric Observatory \citep[SOHO;][]{domingo_1995_soho}, with field-of-views reaching 6~R$_{\odot}$ and 32~R$_{\odot}$, respectively. From the second vantage point, we use the COR1 and COR2 coronagraphs of the Sun Earth Connection Coronal and Heliospheric Investigation \citep[SECCHI;][]{howard_secchi} onboard the Solar TErrestrial RElations Observatory Ahead spacecraft \citep[STEREO-A;][]{kaiser_2008_stereo}, with field-of-views reaching 4~R$_{\odot}$ and 15~R$_{\odot}$, respectively. In addition, we examine EUV observations from the STEREO-A/SECCHI Extreme Ultraviolet imager (EUVI), the Atmospheric Imaging Assembly \citep[AIA;][]{lemen_aia} onboard the Solar Dynamics Observatory \citep[SDO;][]{pesnell_2012_sdo}, and the Extreme Ultraviolet Imager \citep[EUI;][]{rochus_eui} onboard SolO \citep{muller_2013_solo, mullersolo_2020}. However, no EUV data from the latter is available for the time period in question.

We identify a single CME corresponding to the time/plane-of-sky conditions posed by the pulsar observations. The CME is observed in LASCO/C2 and C3, as well as STEREO-A COR1 and COR2. The CME enters the LASCO field of view around 06:12~UT on August 20, 2021 and is measured up to around 11:42~UT according to the SOHO/LASCO CME catalog\footnote{\url{https://cdaw.gsfc.nasa.gov/CME_list/}} \citep{gopalswamy_catalog}, after which CME signatures are still visible in the LASCO field of view, but the leading edge is too faint to be tracked reliably. Just prior to the appearance of the CME in C2, eruptive signatures are observed at the east limb in SDO/AIA 193 \AA{} wavelength, at around 20$^\circ$ south of the equator. Accordingly, a flare is observed in STEREO-A EUVI near the east limb.

We perform a 3D reconstruction of the CME based on white light observations using the Graduated Cylindrical Shell (GCS) model by \cite{Thernisien_2006}. GCS is used to represent the 3D geometry of the flux rope structure of the CME by approximating it with a self-similarly expanding hollow croissant originating from the center of the Sun, where the legs are conical with a circular cross-section and a pseudo-circular front. The empty croissant is defined through 6 parameters which fully define the position, orientation and size of the flux rope and are obtained by fitting its projected 2D features to stereoscopic observations. We perform the reconstruction at the time of optimal visibility of the CME in LASCO and STEREO-A at 08:38~UT, and obtain the following set of GCS parameters: Stonyhurst (also known as Heliocentric Earth Equatorial, HEEQ) longitude 265$^\circ$, latitude $-16$$^\circ$, tilt 70$^\circ$, height 12 R$_{\odot}$, ratio (defines thickness of the croissant legs) 0.32 and half angle (defines opening angle of the croissant) 22$^\circ$. We repeat the reconstruction at several time steps, assuming self-similar expansion (i.e. keeping the ratio and half angle fixed) and radial motion (keeping longitude/latitude fixed). We thus obtain the height-time plot from  which a linear speed of 770~km~s$^{-1}$ is estimated. 

This event is also reported in the Space Weather Database Of Notifications, Knowledge, Information (DONKI) under ID \textit{2021-08-20T06:12:00-CME-001}\footnote{\url{https://kauai.ccmc.gsfc.nasa.gov/DONKI/view/CME/17540/1}}. The catalog entry reports a longitude of -136$^\circ$~($\hat{=}224^\circ$), a latitude of -22$^\circ$, a half-width of 29$^\circ$, and an initial speed of 986~km~s$^{-1}$. Recent comparative studies have demonstrated that independent 3D CME reconstructions can differ substantially in their derived parameters; therefore, discrepancies of this magnitude are not unexpected \citep[e.g.,][]{kay_2024_collectioncollationcomparison}.

At the time of pulsar measurements the CME is nearly exiting the LASCO field of view and quite faint. We therefore forward model the GCS reconstruction to the time of pulsar measurements to confirm that the pulsar is along the LoS which goes through the CME during the time of measurement. In order to achieve this, we assume that the front of the croissant is propagated according to the drag-based model \citep[DBM][]{vrsnak2013propagation} and that the croissant expands self-similarly. This is shown in Figure~\ref{fig:CME-ICME}c).

In addition, we estimate the extent of the CME in the solar equatorial plane, based on the GCS parameters \citep[as described in][]{dumbovic19} and find that, based on the direction and extent, the CME will reach SolO and make a notable impact. The spacecraft configuration at the time of CME detection, as well as the direction and extent of the CME as derived from the GCS reconstruction, are shown in Figure~\ref{fig:CME-ICME}b). We note that looking at the CMEs detected in white light around this time, no other CMEs are directed towards SolO.

\subsection{\textit{In situ} observations of the ICME}\label{sec:insitu}

Along with the remote white light observations, the ICME was observed \textit{in situ} by SolO. At the time of observation, SolO was located at a distance of 0.65~au from the Sun, at an HEEQ longitude of $-79.6^{\circ}$ and latitude of $-1.2^{\circ}$. Thus, SolO remained close to the ecliptic plane during the ICME encounter, as did the LoS to PSR~J1022+1001.

Figure~\ref{fig:CME-ICME}d) presents the magnetometer \citep[MAG;][]{horbury_2020_mag} and Solar Wind Analyser \citep[SWA;][]{owen_2020_swa} measurements obtained by SolO. The CME arrived at SolO at 07:49 UT on August 22, 2021, marked by an increase in total magnetic field strength, plasma density, and temperature, indicating the onset of the sheath region (highlighted in green in Figure~\ref{fig:CME-ICME}d)). A high-speed stream was present ahead of the CME, visible as a region of elevated speed and temperature and reduced density. Consequently, only a modest but sharp increase in solar wind speed is observed at the sheath boundary, suggesting the possible presence of a weak shock. After the sheath region, the flux rope signature begins at 13:38~UT, where a clear rotation of the magnetic field is visible as the shaded red in Figure~\ref{fig:CME-ICME}d), with components given in local spacecraft Radial-Tangential-Normal (RTN) coordinates.

Following the classification scheme introduced by \cite{bothmer_1998} and \cite{mulligan_1998}, magnetic flux ropes are categorized according to their handedness and the orientation of their central axis relative to the ecliptic plane. The three-letter notation (e.g., SWN, ESW) describes the dominant magnetic field direction at the leading edge, center, and trailing edge of the structure as observed in situ. Low-inclination flux ropes have their axis approximately parallel to the ecliptic plane and exhibit a sign change in the normal magnetic field component $B_N$ during the crossing, whereas high-inclination flux ropes have an axis nearly perpendicular to the ecliptic and instead show a sign change in the tangential component $B_T$.

In this event, the normal component $B_N$ changes sign from negative to positive during the passage of the flux rope, while the tangential component $B_T$ remains predominantly positive. This behavior is consistent with a right-handed SWN configuration. Physically, this corresponds to a flux rope with relatively low inclination to the ecliptic plane, an axis oriented toward the west, and a poloidal field directed southward at the leading edge.

We note that the orientation inferred from the remote-sensing reconstruction suggests a higher inclination. However, coronagraph observations show progressive distortion of the CME front, which may indicate rotation of the flux rope. In addition, the tilt angle is typically the least constrained parameter in GCS reconstructions \citep{verbeke23}. Moreover, several studies suggest that flux rope axes may evolve toward lower inclinations as CMEs propagate outward, aligning with the heliospheric current sheet \citep{yurchyshyn2008,isavnin2014three}. Multi-spacecraft \textit{in situ} observations have likewise reported small decreases in flux rope axis inclination with increasing heliocentric distance \citep{good2019,davies2022multi}.

The ICME has completely swept over SolO at around 03:29 UT on August 23, 2021, indicated by the end of the smooth magnetic field rotation. This transition is also marked by an increase in plasma density and a sharp rise in the plasma beta parameter.

%%%%%%%%%%%%%%%%%%%%%%%%%%%%%%%%%%%%%%%%%%%%%%%%%%%%%%%%%%%%%%
\section{Methods}\label{sec:methods}

\subsection{Dispersion and Faraday rotation}

Dispersion and Faraday rotation arise from the propagation of radio waves through a magneto-ionic medium and provide diagnostics of the electron density and magnetic field along the LoS. Dispersion introduces a frequency-dependent delay as the refractive index of an ionized medium depends on frequency. In particular, this delay $\Delta t$ between two radiation frequencies $f_1$ and $f_2$ is expressed as

\begin{equation}\label{eq:dm}
    \Delta t = \mathcal{D}\; {\text{DM}} \left ( \frac{1}{f_1^2} - \frac{1}{f_2^2} \right ),
\end{equation}

with $\mathcal{D}$ being the \textit{dispersion constant} $e^2/(2\pi m_e c)$, with $e$ being the electric charge of an electron, $m_e$ the mass of the electron, and $c$ the speed of light. DM represents the integrated electron density along the propagation path,

\begin{equation}
    \text{DM} = \int_0^Ln_e dl,
\end{equation}

\noindent where $L$ is the distance to the pulsar and $n_e$ is the electron density.

Faraday rotation is the measure of the rotation of the plane of linear polarization $\Psi$,  which is inversely proportional to the square of the frequency. The rotation $\Delta \Psi$ observed at a wavelength $\lambda$ can be written as

\begin{equation}\label{eq:rm}
    \Delta \Psi = \text{RM} \times {\lambda^2},
\end{equation}

\noindent where $\text{RM}$ is defined as

\begin{equation}
    \text{RM} = \frac{e^3}{2 \pi m_e^2 c^4} \int_0^L n_e B_{\parallel} dl,
\end{equation}

\noindent with $B_{\parallel}$ denoting the component of the magnetic field parallel to the LoS. Thus, by calculating both the DM and RM parameters for a specific pulsar observation, it is possible to estimate the average magnetic field component along the LoS as

\begin{equation}
    \langle B_{\parallel}\rangle = 1.23  \frac{\text{RM}(rad\cdot m^{-2})}{\text{DM}(cm^{-3} \cdot pc)}~\mu G.
\end{equation}

Equations (\ref{eq:dm}) and (\ref{eq:rm}) show that, for a fixed precision on measured $\Delta t$ and $\Delta \Psi$, the precision on DM and RM scales as $f^{2}$. Radio observations at low frequencies are therefore strongly favored, motivating the use of instruments such as NenuFAR (10--85 MHz) and LOFAR (110--190 MHz), which operate at the lowest frequencies accessible from the ground. \footnote{This advantage applies at the heliocentric distance considered here. At substantially smaller elongations, however, higher observing frequencies may become preferable because of increased solar contamination and propagation effects, while the intrinsically larger Faraday rotation closer to the Sun partly compensates for the weaker $\lambda^{2}$ dependence.}
\begin{figure}
\includegraphics[width=\linewidth]{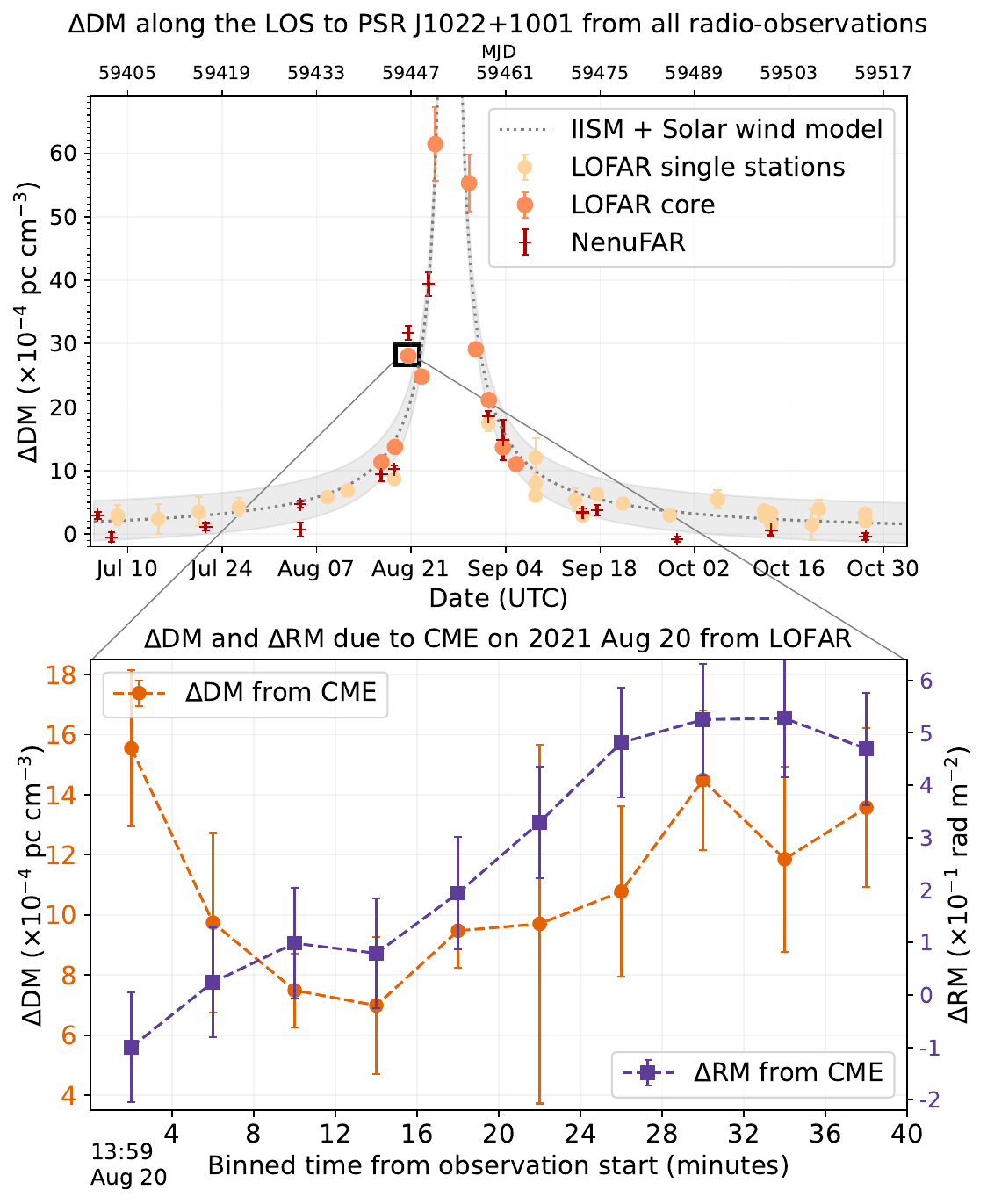}
\caption{Top panel: DM time-series for PSR~J1022+1001 during the 2021 solar conjunction. Bottom panel: DM and RM variations measured from the LOFAR-core observation (started at 13:59 UTC) during the CME occultation to the pulsar LoS. Note that the CME arrives at J1022+1001's projected position at $\sim12:42$ UTC. The 40 minute observation was divided into 4 minute sub-integrations to obtain each point in the lower plot.}
\label{fig:DM_timeserie}
\end{figure}

\subsection{Data analysis of radio data and results}\label{sec:dataanalysis}

In this section, we describe how we derive dispersion measure (DM) and rotation measure (RM) from the LOFAR and NenuFAR observations of PSR~J1022+1001. We then present the time-resolved DM and RM associated with the CME occultation on August 20, 2021.

First, radio-frequency interference (RFI) was removed from both LOFAR and NenuFAR data using the \textit{surgical} mode of the \textsc{coastguard} software suite \citep{lazarus_2016}. LOFAR data are additionally beam-calibrated using the \textsc{dreambeam} software suite \citep{noutsos2015pulsar}, and integrated to a time-resolution of 60~s.

\subsubsection{Estimating the background DM time series}
To calculate a DM value per each observation, we use the technique of \textit{pulsar timing} \citep{lorimer2005handbook}. In this approach, the predicted times-of-arrival (ToAs) of the pulsar's impulses are calculated using a \textit{pulsar timing model} based on an ensemble of parameters which includes the DM. These parameters are estimated via a generalized least-square fit against the TOAs, leaving behind a set of timing residuals. Specifically, if the DM of a certain observation is not correctly defined, we expect a quadratic, frequency dependent trend in the timing residuals at each epoch following Equation~\ref{eq:dm}. Hence, to calculate a DM value per each observation, it is fundamental to have frequency-resolved ToAs for every archive, one per each frequency channel. 

To calculate frequency-resolved ToAs we proceed as follows. At first, we integrate the observations' frequency resolution from 400 frequency channels to 10 in the case of LOFAR, and from 384 to 8 in the case of NenuFAR. This is because the signal-to-noise (S/N) ratio in each channel is not sufficient to derive a precise ToA otherwise. Then, we build a reference template, i.e., a high S/N, analytic representation of the pulse shape per each frequency channel. To achieve this, we stack together all the observations (excluding the ones obtained during the solar passage) of the campaign obtained with one of the telescopes, and we analytically model the result of each frequency channel to remove the noise fluctuations. The template is then cross-correlated with each observation collected with that specific telescope to calculate frequency-resolved ToAs relative to the template position. At this point, the ToAs of each observation are independently fit for DM, in order to obtain DM time series per both LOFAR and NenuFAR. For more details on this procedure (see \cite{donner2019first,donner2020dispersion,Tiburzi_2019,Tiburzi_2021}). The results are shown in Figure~\ref{fig:DM_timeserie}. The uncertainty on each DM measurement is taken as the formal \(1\sigma\) uncertainty returned by the timing fit to the frequency-resolved ToAs. The comparatively large DM uncertainty in the sub-integration around 22~min results in Figure~\ref{fig:DM_timeserie} is due to its lower effective signal-to-noise ratio, which reduces the precision of the timing fit. Since RM is determined independently from the weighted fit to Stokes $Q$ and $U$, its uncertainty does not necessarily follow the same variation as the DM uncertainty.

\subsubsection{DM measurement of CME observations}
 On 20 August 2021, both  LOFAR and NenuFAR observations show an anomalous excess DM of  $\sim1.49\times10^{-3}$ $\pm$0.05$\times$10$^{-3}$ pc~cm$^{-3}$ and $\sim1.5\cdot10^{-3}$$\pm$0.1$\times$10$^{-3}$ pc~cm$^{-3}$, respectively, as shown in the upper panel of Figure~\ref{fig:DM_timeserie}. This excess is observed relative to the IISM+Solar wind model fit (gray bounded region in Figure~\ref{fig:DM_timeserie}) to the whole set of observations acquired during the 2021 solar conjunction campaign.

\begin{figure}
    \centering
    \includegraphics[width=\linewidth, trim = 10mm 11mm 5mm 12mm, clip]{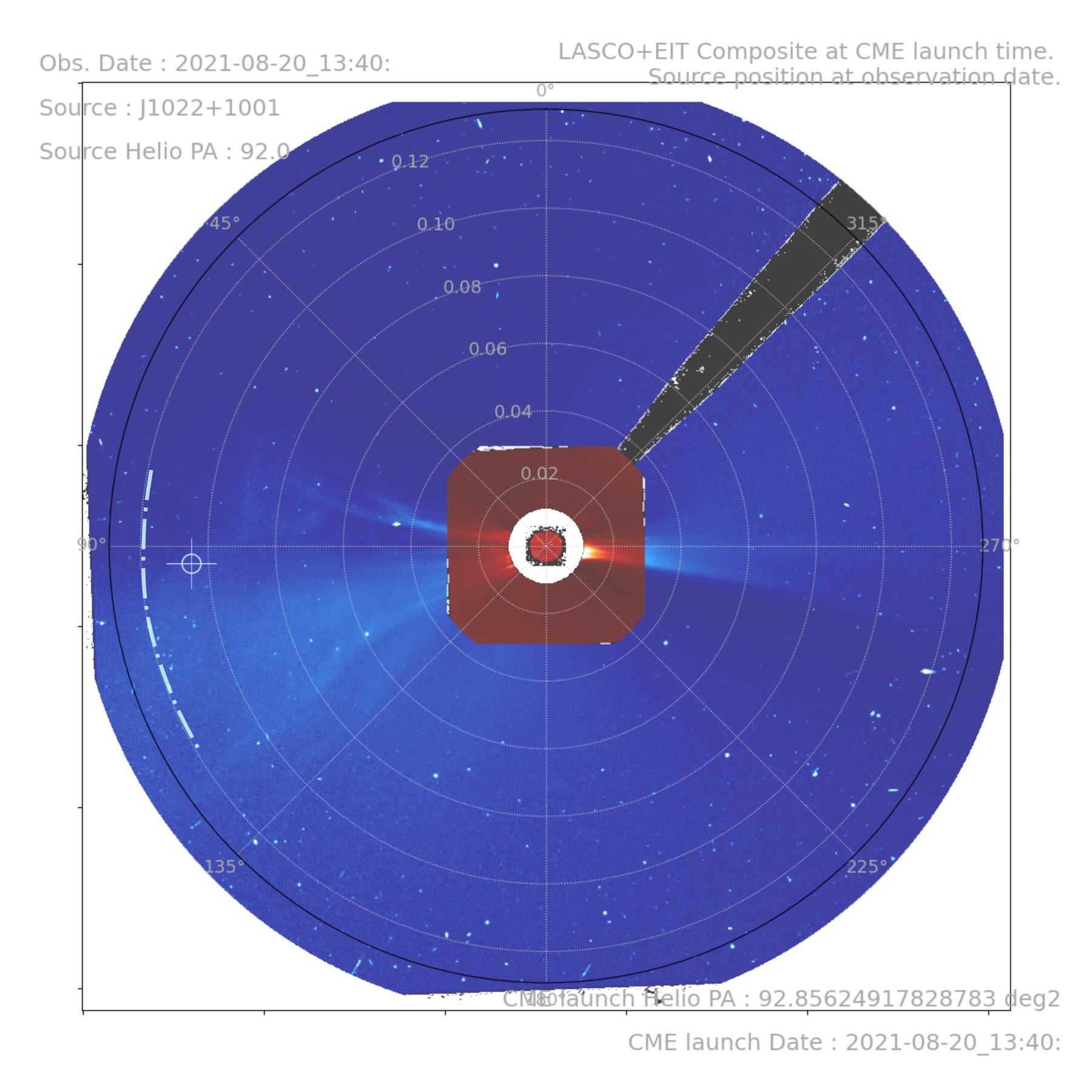}
    \caption{\textsc{CMEchaser} plot showing the position of J1022+1001 with a white crosshair symbol at 13:40~UT on August 20, 2021. The background image shows both the LASCO C3 and C2 images of the event.\label{fig:cmechaser} The dash-dotted line indicates the approximate location of the CME front.}
\end{figure}

At that time the pulsar was at a projected heliocentric distance of $24.6\,R_\odot$, which makes transient solar structures a plausible origin of the excess DM.  This was cross-checked with the \textsc{CMEchaser} package \citep[][see Figure~\ref{fig:cmechaser}]{shaifullahcmechaser_2020}, which used the SOHO LASCO CME Catalog \citep[][ henceforth CDAW Catalog]{gopalswamy_catalog} to confirm that a massive CME launched from the Sun on 20 August 2021 with a position angle of $107^{\circ}$ and a speed of $774$~km~s$^{-1}$, and occulted PSR~J1022+1001. Assuming the CME properties as provided by the CDAW catalog stayed unchanged, it arrived at the pulsar's projected position at $\sim$12:42~UT, and occulted the pulsar throughout the duration of both of our observations. The CME signature is evident in the top panel of Figure~\ref{fig:DM_timeserie} where the excess electron content of the CME increases the observed total DM along the line of sight relative to the background. 

The relatively long integration time and high S/N in the LOFAR observation allowed us to further subdivide it into 4-minute segments to investigate whether DM variations due to the CME transit were detectable. The lower panel of Figure~\ref{fig:DM_timeserie} indicates an initial decrease in DM over the first 16 minutes, followed by a gradual increase for the remainder of the observation. At the start of the LOFAR observation, the pulsar LoS was already occulted by the CME. The remote-sensing images suggest that the LoS subsequently moved from the relatively dense CME front into a lower-density region, before sampling denser material further within the CME. This sequence provides a natural qualitative explanation for the initial decrease and subsequent increase in DM. We refrain from assigning individual DM features to specific internal flux-rope substructures, since the DM measurement constrains only the integrated electron column density along the LoS.

\subsubsection{RM time series}
The RM values were calculated for the LOFAR observation only, because of the insufficient S/N in linear polarization of the NenuFAR observation. Moreover, the LOFAR observation was beam-calibrated, hence enhancing the possibility of linear polarization to be correctly detected. 
To compute the RM per each observation, we first used the \textsc{rmfit} algorithm from the \textsc{psrchive} software suite, that explores a range of RM values to maximize the linearly polarized flux of the input archive. However, we then used a custom Python code from the \textit{NenuPlot}\footnote{\textit{NenuPlot}: \url{https://github.com/louisbondonneau/NenuPlot}} library, chosen particularly for its robust error determination. In this approach, the code starts with the identification of the RM that maximizes the on-pulse linear polarization and then performs a dedicated frequency‐dependent least-squares fit of Stokes $Q$ and $U$ (via \texttt{rmfit}) to derive the final RM value. This parametric fit provides a covariance matrix for the RM, from which the formal error on RM is obtained. The least-squares fit is weighted by the measurement uncertainties in Stokes \(Q\) and \(U\). The covariance-matrix uncertainty therefore reflects the measurement noise in the polarization data, rather than only the goodness of the fit. We thus measured the RM of all the LOFAR observations taken during the 2021 solar conjunction, and for all the observations were able to fit a single Gaussian except for the observation taken during the CME-pulsar occultation, whose RM spectrum appears heavily non-Gaussian. This might show that a single RM value cannot be attributed to the 40-minute observation taken with LOFAR, likely because of the CME transit. 

\begin{figure}
\centering
\includegraphics[width=1\columnwidth]{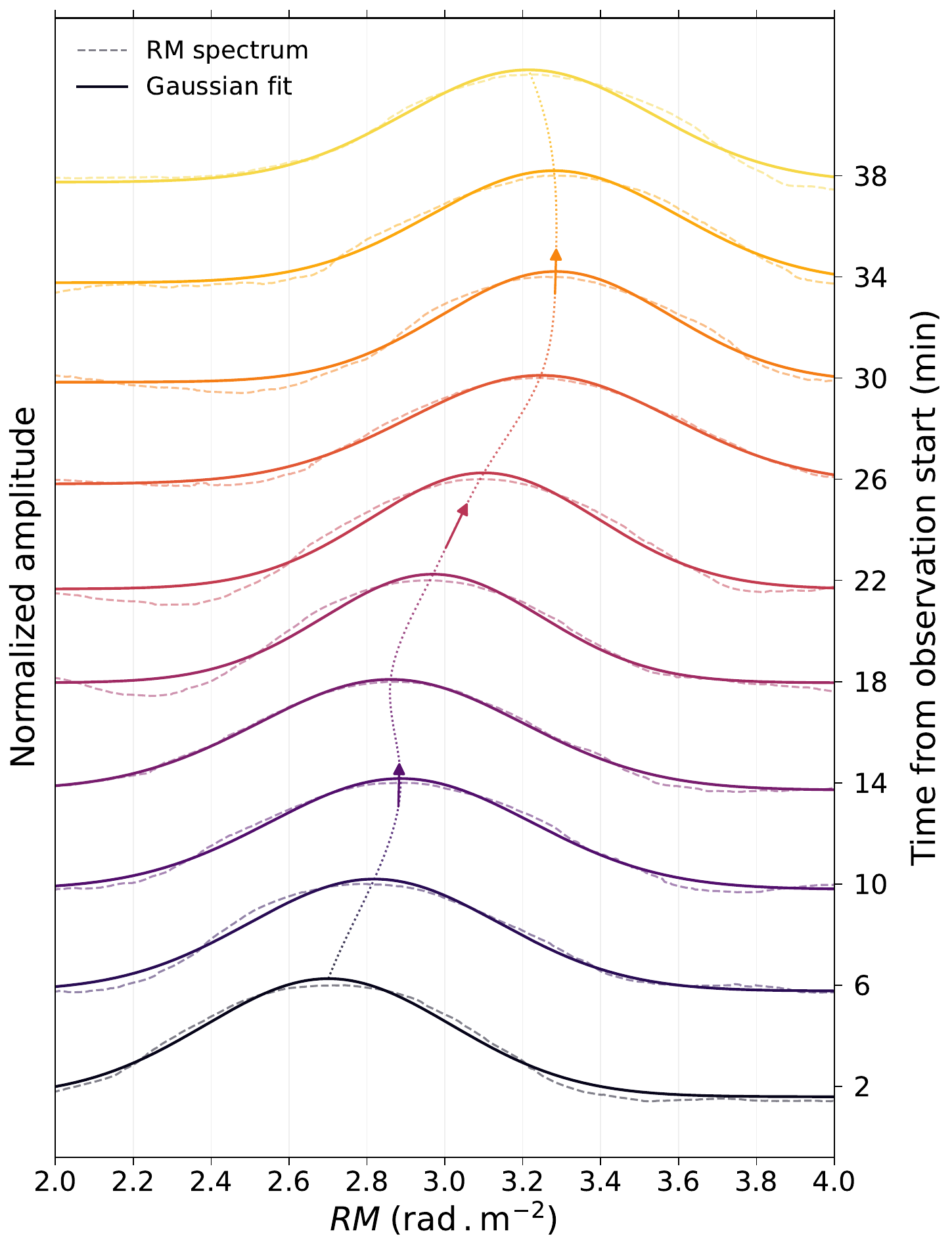}
\caption{The LOFAR rotation measure (RM) spectra for the 4-minute sub-integrations. Dashed curves show the normalized RM spectra of the full LoS, vertically offset for clarity, and solid curves show the corresponding Gaussian fits. Each spectrum is assigned to the midpoint of its sub-integration, indicated by the ticks on the right-hand time axis in minutes from the start of the observation. The dotted line tracks the peaks of the Gaussian fits, with arrows indicating the direction of increasing time.}
\label{fig:rmspectrum} \vspace*{-1\baselineskip}
\end{figure}

To confirm this, we proceeded to break the observation into 4-minute segments, as previously done during the DM analysis, and we calculated an RM value for each segment. We tested different processing choices to improve the signal-to-noise ratio, including integration over pulse-phase bins. These choices produced small shifts in the absolute RM level, while the peak-to-peak RM variation remained consistent. Figure~\ref{fig:rmspectrum} shows the spectra from the final processing used in the analysis. Note that the RM evolution shown in Figure~\ref{fig:rmspectrum} is unique to this observation and is absent in the rest of the J1022+1001 observations for 2021. The results, reported in the lower panel of Figure~\ref{fig:DM_timeserie}, show a variable RM time series during the CME transit that mirrors the spatial changes of the magnetic field inside the CME itself.

Galactic or heliospheric RM calculations are usually compromised by the ionospheric contributions. We correct the observed RM for ionospheric effects using the method described by \citet{porayko_ionosphericRM}, which applies a thin-layer ionospheric model incorporating Global Positioning System–based electron density maps and semi-empirical geomagnetic field models. This approach typically achieves residual uncertainties of approximately 0.06 to 0.07 rad m$^{-2}$ on annual timescales. In the following, we propagate these uncertainties into the extracted RM measurements.

To assess the validity of the measured heliospheric background contribution, we first approximated its order of magnitude. While this simulation output is not directly used to infer the CME contribution to RM and DM from observations, it can confirm whether a simplified model is sufficient in this case. For this, we adopted the Parker spiral model for the heliospheric magnetic field:

\begin{align}
    B_r(\mathbf{r},\theta,\phi) &= B(\theta,\phi_0) \left(\frac{b}{r}\right)^2, \\
    B_\theta(\mathbf{r},\theta,\phi) &= 0, \\
    B_\phi(\mathbf{r},\theta,\phi) &= B(\theta,\phi_0) \left(\frac{\omega}{v_m}\right)(r - b) \left(\frac{b}{r}\right)^2\sin(\theta),
\end{align}

\noindent where $r$ is the radial distance from the Sun, $b$ is the radial distance where the spiral originates, $\theta$ is the polar angle, $\omega$ is the Sun's angular velocity, and $v_m$ is the plasma velocity \citep{1958ApJ...128..664P}. Given that PSR~J1022+1001 lies near the ecliptic plane ($\text{latitude} = -0.06^\circ$), we assume equatorial solar wind conditions and a constant surface magnetic field, $B(\theta = \frac{\pi}{2}, \phi_0) = B_{\text{Surface}}$. This simplifies the field components:

\begin{align}
    B_r(\mathbf{r},\phi) &= B_{\text{Surface}} \left(\frac{R_{\odot}}{r}\right)^2, \\
    B_\phi(\mathbf{r},\phi) &= B_r(\mathbf{r},\phi) \left(\frac{\omega}{v_m}\right)(r - R_{\odot}).
\end{align}

The electron density is modeled using a spherically symmetric solar wind profile. At these heliocentric distances, this model is an effective approximation for such estimations \citep{Tiburzi_2019}:

\begin{align}
    N_e(r) = A_{\text{sw}} \left(\frac{1}{r}\right)^2,
\end{align}

\noindent where $A_{\text{sw}}$ is the electron density of the solar wind at the Earth's orbit.

We discretized the LoS into $N = 3000$ points and computed $B_r$, $B_\phi$, and $N_e$ in astronomical units to directly estimate $\Delta \text{DM}$ and $\Delta \text{RM}$. The parameters used were: $R_{\odot} = 4.65 \times 10^{-3}~\text{au}$, $B_{\text{Surface}} = 4.6 \times 10^{-3}~\text{T}$, $\omega = 2.74 \times 10^{-6}~\text{rad}\cdot\text{s}^{-1}$, $v_m = 2.67 \times 10^{-6}~\text{AU}\cdot \text{s}^{-1}$ (corresponding to a $400$~km~s$^{-1}$ solar wind speed), and $A_{\text{sw}}=7.9~\text{particle}\cdot\text{cm}^{-3}$ at 1~au, measured using 25 pulsars for 2021 \citep{eptadr2}.

We computed the local LoS magnetic field component, $B_\parallel$, and evaluated the integrals of $N_e$ and $N_e B_\parallel$ along the LoS using Simpson's rule, with the appropriate unit conversions. The resulting
heliospheric DM is of order $10^{-3}~\mathrm{pc\,cm^{-3}}$, consistent with previous results \citep{Tiburzi_2021}. With the adopted parameters, the simplified Parker-spiral model gives a heliospheric RM of approximately $3.66~\mathrm{rad\,m^{-2}}$, approximately 2.7 times the empirically determined background RM discussed in Section~\ref{diss:pulsar_B}. The adopted kinematic Parker-spiral approximation does not self-consistently enforce angular-momentum conservation \citep[see][]{1967ApJ...148..217W}. We emphasize that this calculation is used only as an order-of-magnitude consistency check and \textit{is not used to determine the background RM subtracted from the CME observation.}

\subsection{CME modelling}\label{sec:modelling}
The semi-empirical 3DCORE model for CME flux ropes \citep{Mostl2018,Weiss2021a, Weiss2021b}, most recently described in \cite{Rudisser2024}, was employed to generate a 3D reconstruction of the CME under study. The reconstruction is based on \textit{in situ} magnetic field measurements obtained by SolO. Specifically, the model is fitted to the magnetic field observations at SolO, and the derived parameters are then used to evaluate the magnetic field at arbitrary positions and times within the heliosphere.

The 3DCORE model describes a toroidal flux rope with a Gold–Hoyle-like magnetic field and an elliptical cross-section. The structure expands self-similarly as it propagates outward while remaining magnetically connected to the Sun at both ends. The outward propagation follows a drag-based approach \citep{vrsnak2013propagation}. By fitting to the \textit{in situ} magnetic field data, an ensemble of model solutions is obtained using a Monte Carlo–based algorithm that samples the full parameter space \citep{Weiss2021a}. Each independent set of fitted parameters constitutes one ensemble member, i.e., one possible three-dimensional realization of the CME that is consistent with the \textit{in situ} observations.

To achieve faster convergence, the parameter ranges were restricted prior to fitting. The DONKI catalogue values were used to define sufficiently broad search ranges around the reported longitude, latitude, and initial velocity, but were not imposed as fixed values. This allows the reconstruction to account for known uncertainties in remote-sensing-derived CME parameters \citep{kay_2024_collectioncollationcomparison}, and for the relatively rigid flux-rope geometry assumed by 3DCORE. The CME launch time was fixed to August 20, 2021, 09:44~UT at a heliocentric distance of 21.5~$R_\odot$, again following the DONKI entry. A summary of the input parameter ranges and the resulting fitted values is presented in Table \ref{tab:fitting_params}.

\begin{table}
\centering
\caption{Model parameter ranges and the corresponding results: longitude, latitude, inclination, diameter at 1~au ($D_{1\mathrm{AU}}$), aspect ratio ($\delta$), launch radius ($R_{0}$), launch velocity ($V_{0}$), expansion rate ($n_{a}$), background drag ($\Gamma$), solar wind speed ($V_{\mathrm{sw}}$), twist factor ($T_{f}$), magnetic decay rate ($n_{b}$) and magnetic field strength at 1~au ($B_{1\mathrm{AU}}$). The longitude and latitude of the flux rope apex are given in HEEQ coordinates. Parameters that were held fixed during fitting are shown as a single value (no brackets).}

\begin{tabular}{l c c c}
Parameters & Units & Range / Fixed & Results / Value \\[0.5ex]
\hline
Longitude & deg & $[-176,\,-96]$ & $-104 \pm 5$ \\[0.5ex]
Latitude & deg & $[-52,\,8]$ & $3 \pm 3$ \\[0.5ex]
Inclination & deg & $[0,\,360]$ & $147 \pm 4$ \\[0.5ex]
$D_{1\mathrm{AU}}$ & au & $[0.15,\,0.35]$ & $0.27 \pm 0.02$ \\[0.5ex]
$\delta$ &  & $2$ & $2$ \\[0.5ex]
$R_{0}$ & $R_\odot$ & $21.5$ & $21.5$ \\[0.5ex]
$V_{0}$ & km~s$^{-1}$ & $[736,\,1236]$ & $961 \pm 134$ \\[0.5ex]
$n_{a}$ &  & $1.14$ & $1.14$ \\[0.5ex]
$\Gamma$ &  & $[0.1,\,2.4]$ & $1.8 \pm 0.4$ \\[0.5ex]
$V_{\mathrm{sw}}$ & km~s$^{-1}$ & $[374,\,574]$ & $406 \pm 25$ \\[0.5ex]
$T_{f}$ &  & $[-100,\,100]$ & $41 \pm 11$ \\[0.5ex]
$n_{b}$ &  & $1.64$ & $1.64$ \\[0.5ex]
$B_{1\mathrm{AU}}$ & nT & $[5.0,\,75.0]$ & $13.5 \pm 1.9$ \\[0.5ex]
\hline
\end{tabular}
\label{tab:fitting_params}
\end{table}

\begin{figure}
    \centering
    \includegraphics[width=1\columnwidth]{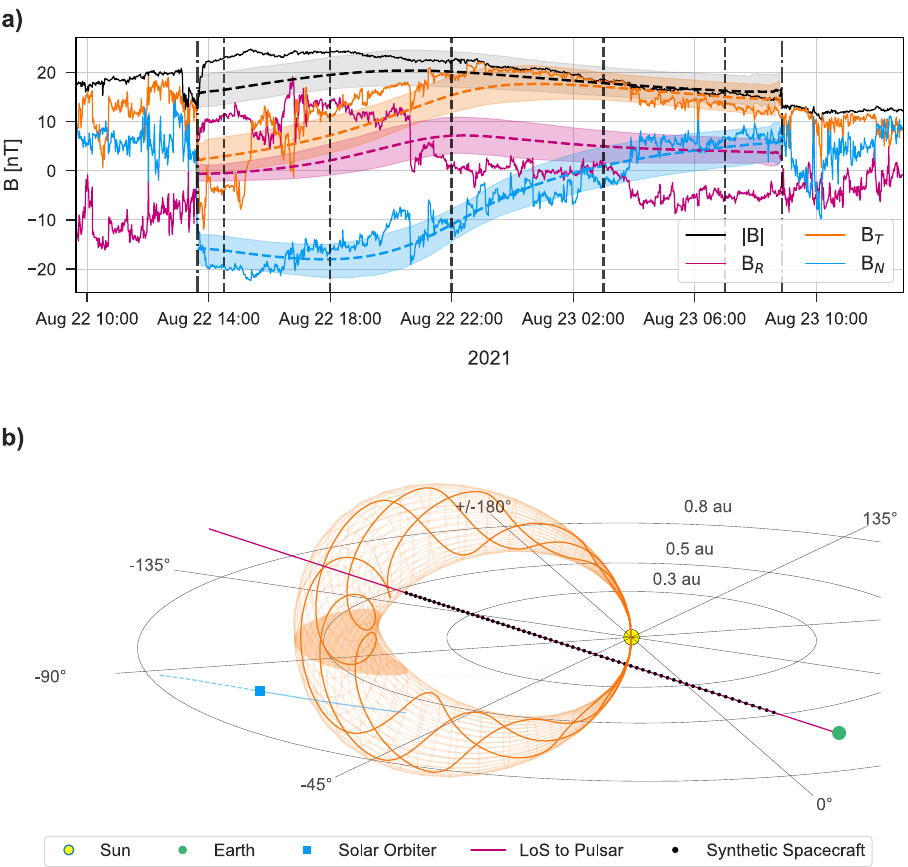}
    \caption{3DCORE fitting results: a) Magnetic field measurements from SolO along with the 3DCORE ensemble fit. Solid lines show the Solar Orbiter measurements, dashed coloured lines show the ensemble mean, and the shaded region represents the $2\sigma$ spread of the ensemble. Vertical black dash-dotted lines indicate the start and end of the time window used for the 3DCORE reconstruction. The grey dashed lines mark the fitting points where the model output is compared to the \textit{in situ} data. The magnetic field data is shown in the local spacecraft Radial-Tangential-Normal (RTN) coordinate system. b) 3D reconstruction of the ensemble mean, together with three example magnetic field lines within the toroidal structure, shown 36~hours after the CME reached 21.5~$r_\odot$. The orange shaded area indicates the intersection of the modeled CME with the ecliptic plane. Also shown are the positions of Earth (green dot) and SolO (blue square), along with its past and future trajectory (dashed and solid lines, respectively). The straight magenta line extending from Earth indicates the LoS to the pulsar, with synthetic spacecraft placed along it (back dots). The pulsar LoS passes behind the central CME structure in this view and does not intersect either CME leg.}
    \label{fig:fitresults_all}
\end{figure}

Figure~\ref{fig:fitresults_all}a) shows the in situ magnetic field measurements from SolO (solid lines) together with the corresponding 3DCORE ensemble reconstruction. The shaded region represents the 2$\sigma$ spread, which can be interpreted as a measure of uncertainty in the reconstruction, while the dashed colored lines indicate the ensemble mean. Vertical black dash-dotted lines mark the start and end of the time interval used for the fitting procedure. Vertical black dashed lines indicate the specific fitting points at which the model output is compared to the \textit{in situ} data. These fitting points were distributed throughout the ICME interval, with slightly increased sampling near the beginning of the event to capture more extreme values. The last fitting point was placed relatively close to the end of the identified ICME interval to ensure that the reconstructed flux rope reproduces the full ICME duration.

The mean normalized root mean square error ($\epsilon = 0.43 \pm 0.01$) indicates a good agreement between the reconstructed ensemble and the \textit{in situ} data at the selected fitting points. The overall trend and magnitude of the total magnetic field $|B|$ are well reproduced. The polarity reversal of $B_R$ however, is inverted in the reconstruction, with its magnitude underestimated at the start and overestimated toward the end of the interval. This behavior is not unexpected, as the radial component is typically the most challenging to fit. For the present analysis, this discrepancy is of limited importance, since the radial component is largely perpendicular to the pulsar LoS. The tangential component, $B_T$, is initially slightly overestimated but aligns well with observed values throughout the remainder of the flux rope. The normal component, $B_N$, is reproduced particularly well. The fitted twist factor $T_f = 41 \pm 11$ (corresponding to approximately 4.3 turns along the full structure) indicates a right-handed flux rope. Together with a fitted inclination of $147 \pm 4^\circ$, the ensemble mean solution corresponds to an SWN-type flux rope.

It is important to emphasize that the Monte Carlo algorithm yields 1024 ensemble members, i.e., 1024 independent model realizations consistent with the \textit{in situ} constraints. While the vast majority of these ensemble members correspond to an SWN configuration, four out of 1024 are classified as ESW, representing a right-handed but high-inclination flux rope. This variation is not unexpected, given that both the tangential component $B_T$ and the normal component $B_N$ exhibit slight bipolarity in the \textit{in situ} data. Since the inferred inclination depends on which component is interpreted as bipolar or unipolar, the classification can shift between low- and high-inclination solutions for a subset of parameter combinations.

Nevertheless, the vast majority of ensemble members favor a low-inclination configuration, and we therefore conclude SWN to be the most probable flux rope type for this event. Importantly, even in the minority high-inclination solutions, the inferred handedness and overall axis orientation remain consistent. The ambiguity thus primarily affects the exact inclination angle rather than the fundamental magnetic structure.

Figure~\ref{fig:fitresults_all}b) illustrates the 3D reconstruction of the ensemble mean identified through the fitting process. The outer boundary of the CME is shown as a grid, containing three example magnetic field lines. The LoS from Earth to the pulsar is indicated as a straight magenta line. Also shown are the positions of Earth and SolO.

%%%%%%%%%%%%%%%%%%%%%%%%%%%%%%%%%%%%%%%%%%%%%%%%%%%%%%%%%%%%%%
\section{Results}\label{sec:results}

\subsection{Pulsar-derived $\langle B_{\parallel}\rangle_{\text{PSR}}$ }\label{diss:pulsar_B}

The derivation of $\langle B_{\parallel}\rangle_{\text{PSR}}$ requires the isolation of the contribution of the CME to both the total DM and total RM measurements. To derive the $\text{DM}_{\text{CME}}$, the solar wind spherical model in \cite{Tiburzi_2019} plus a third-order polynomial for $\text{DM}_{\text{ISM}}$ was chosen as a sufficient model to isolate different DMs. A Bayesian fit of the total model was performed using a Markov Chain Monte Carlo algorithm. This fit estimated a solar amplitude $A_{\text{sw}}= 13.41\pm0.49$ $  \text{particle}\cdot\text{cm}^{-3}$  and a background $\text{DM}_{\text{ISM}}=10.2534\pm4.15\cdot 10^{-5}\text{pc}\cdot\text{cm}^{-3}$  as shown in Figure~\ref{fig:DM_timeserie}. While, the fitted $A_{\text{sw}}$ yields a higher electron density than the standard value (measured using 25 pulsars for 2021 \citep{eptadr2}), this difference can be explained by the presence of a higher density region in the solar wind, that can be observed in remote imaging. By subtracting the background DM from the DM time series, one can isolate $\text{DM}_{\text{CME}}$ (see Figure~\ref{fig:DM_timeserie}). 

% The extraction of the CME's contribution to RM requires a more delicate procedure than the DM one. First, to estimate the background RM (Interstellar + Solar wind),  the CME observation was fortunately conducted between two others, each taken two days apart. These ordinary observations were corrected for the ionospheric RM and used to interpolate the background RM of $1.37\pm0.08\;\text{rad}\cdot\text{m}^{-2}$. This observed background RM matches exactly the estimated heliospheric RM, which arises from the peculiarity of J1022+1001 RM being mainly dominated by heliospheric and ionospheric RM.
The extraction of the CME’s contribution to RM requires a more delicate procedure than the DM one. To estimate the background RM (interstellar medium + solar wind), the CME observation was conducted between two observations obtained two days before and after the event. These observations were corrected for the ionospheric RM and interpolated to the CME epoch, yielding a background RM of \(1.37\pm0.08\,\mathrm{rad\,m^{-2}}\). This empirically determined value is used for the RM background subtraction and is independent of the value of \(A_{\rm SW}\) adopted in the simplified Parker-spiral calculation described in Section 3.2.

The ionosphere-corrected RM changes only from \(1.40\) to \(1.34\ {\rm rad\,m^{-2}}\) between the observations two days before and after the CME, corresponding to a gradual variation of \(\sim0.015\ {\rm rad\,m^{-2}\,day^{-1}}\); this slow evolution is accounted for by interpolating between the two measurements.

Subsequent ionospheric corrections of each 4 min sub-integration using established models ~\citep{porayko_ionosphericRM} allowed the CME RM variations to be clearly identified. 
For each 4-minute sub-integration, the CME contribution is calculated as
\begin{equation}
    {\rm RM}_{\rm CME}={\rm RM}_{\rm obs}-{\rm RM}_{\rm iono}-{\rm RM}_{\rm bg},
\end{equation}
where \({\rm RM}_{\rm bg}=1.37\pm0.08\ {\rm rad\,m^{-2}}\). For example, at MJD 59446.60139, 26 minutes from the start of our observations, the measured total RM is \(3.299\ {\rm rad\,m^{-2}}\) and the modeled ionospheric contribution is \(1.471\ {\rm rad\,m^{-2}}\). Subtracting these together with the interpolated background gives
\begin{equation}
    {\rm RM}_{\rm CME}=3.299-1.471-1.370=0.458\ {\rm rad\,m^{-2}},
\end{equation}
corresponding to the CME-associated RM shown in Figure~\ref{fig:DM_timeserie}. That figure displays the dynamic change of RM due to the CME crossing. Consequently, one can measure the magnetic field along the LoS by combining both $\text{RM}_{\text{CME}}$ and $\text{DM}_{\text{CME}}$ measurements. The averaged magnetic field along the LoS is given by

\begin{align}
    \langle B_{\parallel}\rangle_{\text{PSR}} \equiv\frac{\int_{0}^{d} n_{e}B_{\parallel} dl }{\int_{0}^{d} n_{e} dl }= 1.23 \mu G \left(\frac{RM}{rad\cdot m^{-2}}\right)\left(\frac{DM}{cm^{-3} \cdot pc}\right)^{-1}.
\end{align}
The uncertainty in $\langle B_{\parallel}\rangle_{\rm PSR}$ is obtained by propagating the uncertainties in the CME-associated RM and DM measurements, assuming they are independent:
\begin{equation}
    \sigma_{\langle B_{\parallel}\rangle}
    =
    \left|\langle B_{\parallel}\rangle_{\rm PSR}\right|
    \sqrt{
        \left(\frac{\sigma_{{RM}_{ CME}}}{{ RM}_{\rm CME}}\right)^2
        +
        \left(\frac{\sigma_{{ DM}_{\rm CME}}}{{ DM}_{\rm CME}}\right)^2
    }
    .
\end{equation}
For the RM contribution, the ionospheric uncertainty is included through the fractional term $\sigma_{{RM}_{\rm iono}}/{ RM}_{\rm iono}$, with $\sigma_{{RM}_{\rm iono}}=0.08~{\rm rad\,m^{-2}}$.

The resulting average parallel magnetic field associated with the CME is presented in Figure~\ref{fig:average_Bpar} a). We observe a significant initial rise in $\langle B_{\parallel}\rangle_{\text{PSR}}$  from $-8.65\pm1.67$ nT to a peak value of $63.24\pm19.16$ nT within the first 28 minutes of observation, after which the magnetic field stabilizes, fluctuating between $62.19\pm18.49$ nT and $47.49\pm10.58$ nT over the subsequent 12 minutes.

\subsection{CME modeling estimation of $\langle B_{\parallel}\rangle_{\text{3D}}$  }

Although 3DCORE was not originally designed to compute LoS magnetic field averages, it provides the full three-dimensional magnetic field vector at arbitrary positions and times for each ensemble member. This allows us to evaluate the magnetic field along the pulsar LoS for every model realization.

To sample the magnetic field along the LoS, we introduce a set of synthetic spacecraft, which are fixed spatial sampling points distributed along the LoS at heliocentric distances spanning the expected extent of the CME. Their positions are shown as black dots along the magenta LoS in Figure~\ref{fig:fitresults_all}b).

For each ensemble member (i.e., for each independent set of model parameters), we evaluate the magnetic field vector at the position of every synthetic spacecraft as a function of time. The magnetic field is then projected onto the LoS to obtain the local LoS component $B_{\parallel}(s,t)$, where $s$ is the coordinate along the LoS.

\begin{figure}[htbp]
\includegraphics[width=\columnwidth]{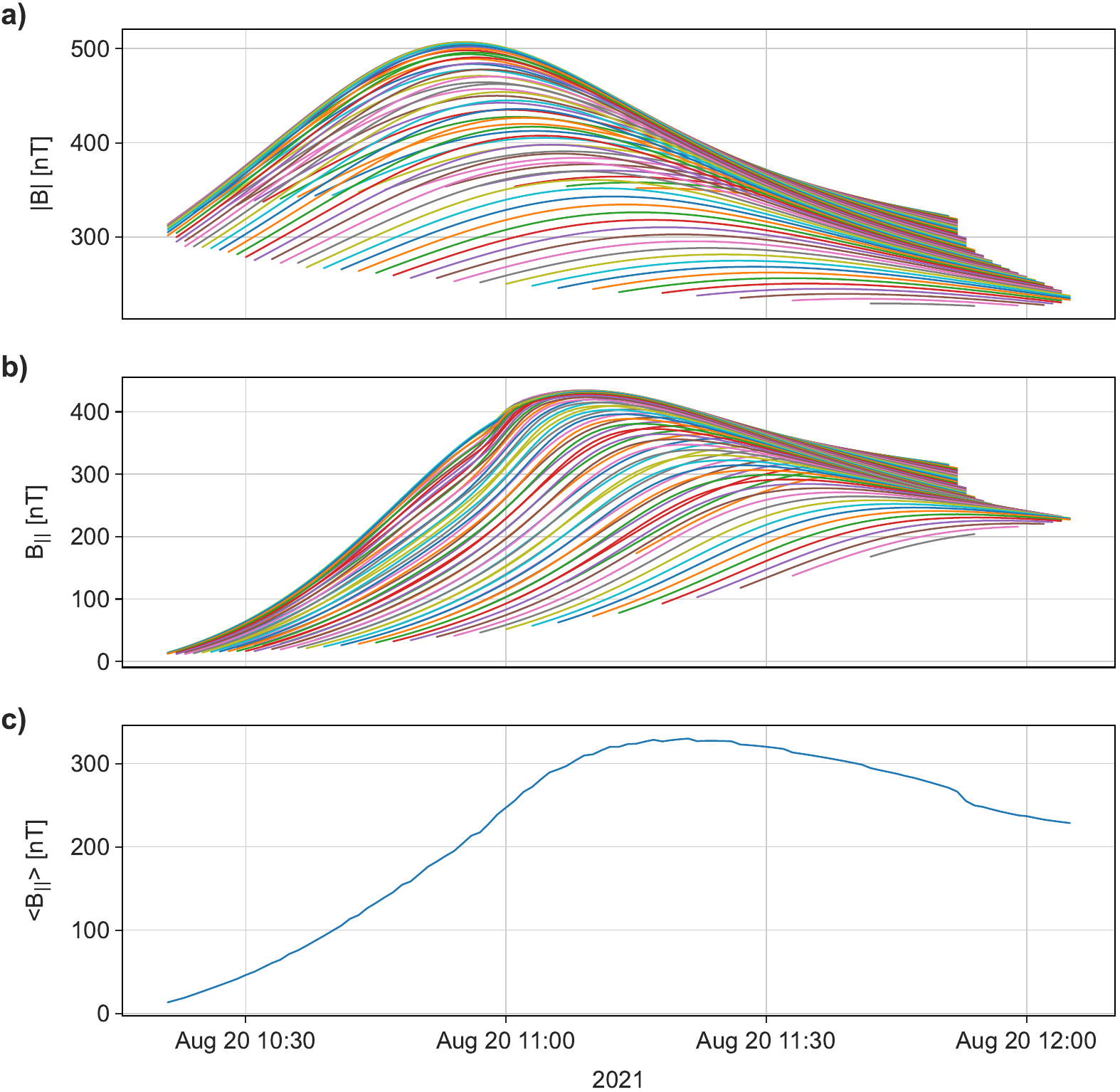}
\caption{a) Total magnetic field as measured by each of the synthetic spacecraft. b) $B_{\parallel,3D}$ as measured by each of the synthetic spacecraft. In panels (a) and (b), each colored curve represents an individual synthetic spacecraft, while the colors are used only to distinguish the sampling points and do not encode their position along the LoS.  c) $\langle B_{\parallel}\rangle_{\text{3D}}$, obtained by integrating and normalizing $\langle B_{\parallel}\rangle_{\text{3D}}$ along the LoS.}
\label{fig:combined_method}
\end{figure}

At each time step, we compute the mean LoS magnetic field for that ensemble member by averaging $B_{\parallel}(s,t)$ over all synthetic spacecraft that are located inside the CME at that time. This yields one time series $\langle B_{\parallel}\rangle(t)$ per ensemble member. Figure~\ref{fig:combined_method} illustrates this procedure for one representative ensemble member: Panel a) shows the total magnetic field magnitude at each synthetic spacecraft, panel b) shows the corresponding projected $B_{\parallel}$ values, and panel c) shows the resulting mean $\langle B_{\parallel}\rangle(t)$ obtained by averaging over all contributing synthetic spacecraft at each time step.

Due to the simplified modeling of both the background solar wind speed and the CME speed, there is considerable shift and variation in the start time of the LoS crossing between ensemble members. To enable a meaningful comparison of the temporal evolution, we align the start times of the $\langle B_{\parallel}\rangle(t)$ profiles before computing the ensemble statistics. After alignment, we calculate the ensemble mean and standard deviation of  $\langle B_{\parallel}\rangle(t)$ across all 1024 ensemble members.

The pulsar observations effectively begin approximately 30 minutes after the initial LoS interception by the CME. For comparison, the observational time axis is therefore shifted relative to the synthetic profiles. Since the pulsar observations terminate before the modeled LoS exit, variations in the end times are not accounted for.

To investigate how different flux rope configurations affect the predicted LoS signature, we repeat the procedure after modifying the twist sign and/or shifting the inclination by $180^{\circ}$, thereby generating synthetic $\langle B_{\parallel}\rangle(t)$ profiles for the four flux rope types considered (SWN, NES, NWS, SEN). The resulting ensemble-averaged profiles, as well as their standard deviation are shown in Figure~\ref{fig:average_Bpar}b).

%%%%%%%%%%%%%%%%%%%%%%%%%%%%%%%%%%%%%%%%%%%%%%%%%%%%%%%%%%%%%%
\section{Discussion}\label{sec:discussion}

By combining pulsar measurements with 3DCORE modeling, we compare the magnetic structure inferred from radio observations with synthetic signatures derived from a three-dimensional reconstruction of the CME. Sampling the model magnetic field along the pulsar LoS allows us to construct synthetic $\langle B_\parallel \rangle_{3D}$  profiles that can be compared with the pulsar-derived $\langle B_\parallel \rangle_{\rm PSR}$ time series. Although these quantities are not strictly equivalent observables (see discussion below), this comparison provides a valuable means of testing the consistency between modeled CME magnetic structure and pulsar observations. The comparison also benefits from the favourable relative geometry of SolO and the pulsar LoS for this event. However, the Faraday-rotation measurement remains sensitive only to the magnetic-field component projected along the pulsar LoS. In particular, the radial component measured by SolO is nearly perpendicular to the pulsar LoS and therefore contributes only weakly to the observed RM. Consequently, the pulsar measurement alone cannot fully constrain this component, even though it provides a strong discriminator between the candidate flux-rope configurations considered here.

The comparison shows that the temporal evolution and polarity of the signal are strongly sensitive to the flux rope configuration. Among the four flux rope types considered (SWN, NES, NWS, SEN), only the SWN configuration, consistent with the \textit{in situ} observations, reproduces the observed sign and overall temporal behavior seen in the pulsar measurements. The synthetic profiles show an increase in $\langle B_\parallel \rangle$ as the LoS enters the CME followed by a gradual decrease while the LoS remains within the structure, consistent with the observed evolution. Although the ensemble-mean SWN profile remains positive throughout the crossing, individual ensemble members can initially reach negative $\langle B_\parallel\rangle_{\rm 3D}$ values. Such behavior is therefore possible within the fitted ensemble, but occurs only for a small subset of members. The remaining configurations produce incompatible polarity or evolution and can therefore be excluded for this event. 

At the same time, the comparison highlights several remaining limitations and mismatches that indicate priorities for future work. First, the synthetic signatures show substantial variability in start times and durations between ensemble members. These shifts arise primarily from simplified assumptions in 3DCORE concerning background solar wind speed and CME propagation. Such timing sensitivities are well known from multipoint CME studies, where small variations in input parameters can lead to large differences in predicted arrival times and durations \citep[e.g.,][]{Weiss2021b, davies2022multi, davies_2024}. In our analysis, aligning the start times enables comparison of the magnetic signature shapes, but the spread underscores that modeling CME evolution close to the Sun with 3DCORE remains challenging.

\begin{figure}
\centering
\includegraphics[width=\columnwidth]{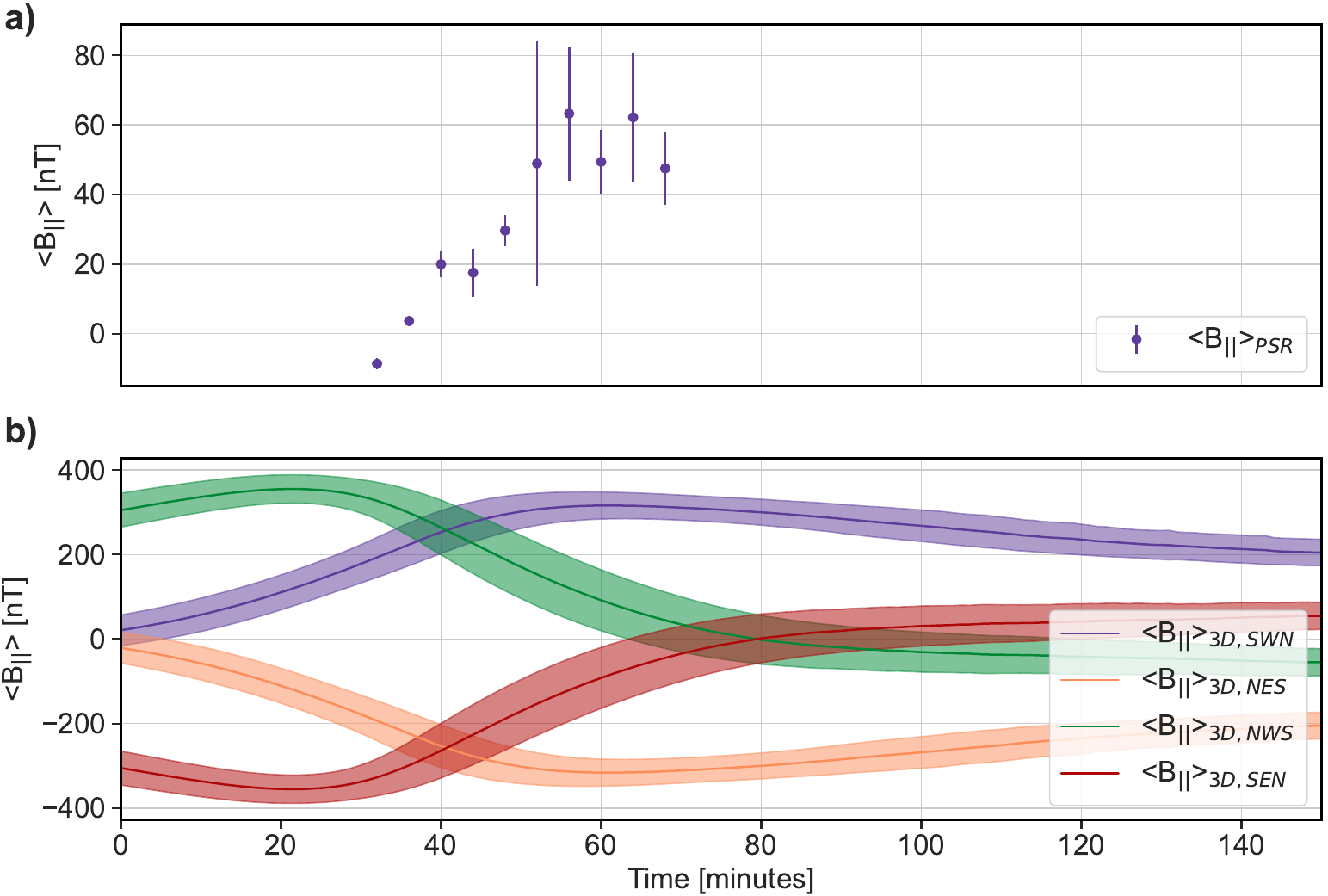}
\caption{(a) $\langle B_{\parallel}\rangle_{\text{PSR}}$ measurements derived through 4 minutes sub-integrations. (b) $\langle B_{\parallel}\rangle_{\text{3D}}$ as modeled by 3DCORE for the original flux rope type (SWN), as well as the three alternative flux rope types (NES, NWS, and SEN). The shaded regions indicate one standard deviation ($1\sigma$) across the ensemble members.}\label{fig:average_Bpar}
\end{figure}

Second, the modeled amplitudes are systematically larger than the pulsar-derived values. While $\langle B_\parallel \rangle_{\rm PSR}$ increases from approximately $-9~\mathrm{nT}$ to a peak near $63~\mathrm{nT}$ (see Figure~\ref{fig:average_Bpar}a)), the SWN ensemble mean shown in Figure~\ref{fig:average_Bpar}b) reaches values of the order of $300~\mathrm{nT}$, roughly a factor of five larger. Differences in absolute amplitude are not unexpected when comparing pulsar-derived quantities with model magnetic fields, since many Faraday rotation studies constrain flux rope orientation primarily through the normalized shape of the signal rather than its absolute magnitude \citep[e.g.,][]{jensen_2008_faradayrotationobservations}. Several factors may contribute to this mismatch. The 3DCORE parameters are constrained by \textit{in situ} measurements at 0.65~au, whereas the LoS intersects the CME at much smaller heliocentric distances, where the simplified assumptions within 3DCORE may not describe the true radial evolution of the magnetic field \citep{mostl2026magneticfieldevolutioninterplanetary}. Nevertheless, observations by Parker Solar Probe (PSP) demonstrate that magnetic field strengths of several hundred nT are not unusual for ICMEs at these heliocentric distances. The projected heliocentric distance of the pulsar LoS, $24.6R_\odot$, corresponds to approximately 0.11~au. ICME encounters observed by PSP at comparable heliocentric distances ($\sim0.07$–$0.14$~au) show maximum magnetic field strengths ranging from approximately $300$~nT to more than $1600$~nT \citep[ICMECAT;][]{moestl_2025_helio4castinterplanetarycoronal,mostl2026magneticfieldevolutioninterplanetary}. The $\sim300$~nT values predicted by 3DCORE are therefore within the range of magnetic field strengths observed close to the Sun.

Additional differences may arise from uncertainties in the magnetic field reconstruction itself. The \textit{in situ} fit to the SolO measurements appears to overestimate the early magnitude of the tangential component $B_T$, which could increase the projected LoS component. Beyond fit-related uncertainties, 3DCORE assumes a simplified, symmetric flux rope geometry with uniform twist. Real CMEs frequently deviate from such idealized configurations due to interactions with the ambient solar wind or neighbouring transients \citep[e.g.,][]{davies2021solo, al-haddad_2025_magneticfieldstructure}. Consequently, the pulsar LoS may sample a region of the CME that differs significantly from the part later encountered \textit{in situ}. In addition, interaction or partial merging with surrounding structures could further modify the local magnetic structure and amplitude \citep[e.g.,][]{scolini_2022_causesconsequencesmagnetic, lugaz_2018_spatialcoherencemagnetic,davies2022multi}. In the present event, however, SolO observes a comparatively clean flux rope signature, and the overall agreement of the 3DCORE fit with \textit{in situ} data suggests that the assumption of an idealized flux rope remains a reasonable approximation.

In parallel, the pulsar-derived and model-derived quantities are not directly equivalent observables, which may be the major contribution to the amplitude discrepancy. The pulsar-derived quantity is obtained from the ratio $\Delta{\rm RM}/\Delta{\rm DM}$, corresponding to a density-weighted LoS average. In contrast, 3DCORE provides only the magnetic field, and the synthetic signatures are constructed by averaging $B_\parallel(s,t)$ at equally spaced sampling points along the LoS within the model CME,

\begin{equation}
\langle B_\parallel \rangle_{\rm 3D}(t)\approx \frac{1}{N_{\rm in}(t)}\sum_{i\in{\rm CME}} B_\parallel(s_i,t).
\end{equation} 

This quantity represents an unweighted geometric average and should therefore be regarded as a proxy describing how the modeled magnetic structure projects onto the LoS, rather than a direct equivalent of the pulsar observable. If density variations exist within the CME, as expected, the pulsar measurement preferentially weights regions of enhanced plasma density, while the model average treats all sampled regions equally. Differences in internal density structure could therefore produce substantial amplitude differences even if the underlying magnetic structure is reproduced correctly. Incorporating realistic density estimates into forward modeling will be essential for future quantitative comparisons \citep[e.g.,][]{wood_inferences_2020}.

Uncertainties also arise in deriving the pulsar-based measurements. The dispersion measure, as described in Equation~\ref{eq:dm}, assumes a cold plasma where the cyclotron frequency is negligible compared to the frequency of the observed radio-wave. However, this assumption may be violated in the CME environment. Hence, the CME requires a more adapted and modified DM~\citep{li_2019MNRAS.484.5723L} to yield a better $\langle B_\parallel \rangle_{\rm PSR}$ measurement. Additionally, the extraction of CME contributions to RM and DM requires subtraction of background interstellar and solar wind components, and uncertainties in these background estimates propagate directly into $\langle B_\parallel \rangle_{\rm PSR}$. Although the background RM and DM are constrained using nearby observations and empirical solar wind models, small systematic errors or unrelated temporal variations may influence the inferred amplitudes.  

Despite the amplitude differences, the agreement in sign and temporal evolution between the observed and modeled $\langle B_\parallel \rangle$ profiles indicates that pulsar measurements capture important information about the large-scale magnetic structure of the CME. This demonstrates the potential of monitoring millisecond pulsars (MSPs) as probes of global heliospheric magnetic structure \citep{Tiburzi_2019}. In this context, a single pulsar acts as a remote magnetometer that is sensitive to large-scale magnetic field structure and can distinguish between incompatible flux rope scenarios. A single pulsar LoS does not, by itself, uniquely determine the full three-dimensional magnetic-field geometry of a CME, since it constrains only the density-weighted magnetic-field component projected along that LoS. Its strength therefore lies in discriminating between candidate magnetic configurations when combined with an independently constrained CME morphology. This is directly analogous to the analysis of \citet{wood_inferences_2020}, particularly their Figure~10(b-e), where even a single-LoS Faraday-rotation measurement was able to exclude a substantial range of otherwise plausible flux-rope configurations.

Future progress will require forward modeling that reproduces the pulsar observables directly, including density structure and LoS weighting, as well as more realistic deformable CME models \citep{weiss_2024_distortedmagneticflux} and systematic exploration of interception geometries. Integrating pulsar constraints into CME reconstruction methods, for example by extending 3DCORE fitting to include LoS signatures directly, represents a promising next step. Together with repeated pulsar conjunction observations and upcoming missions and instruments (e.g., MOST, \citeauthor{gopalswamy_multiview_2024} \citeyear{gopalswamy_multiview_2024}; and FETCH, \citeauthor{jensen_2023_faradayeffecttracker} \citeyear{jensen_2023_faradayeffecttracker}) providing additional viewpoints and measurements, this approach offers promising prospects for improving CME reconstruction and, ultimately, space weather forecasting.

%%%%%%%%%%%%%%%%%%%%%%%%%%%%%%%%%%%%%%%%%%%%%%%%%%%%%%%%%%%%%%
\section{Conclusions}\label{sec:conclusions}

Through dedicated monitoring campaigns within the LOFAR and NenuFAR \citep{ZarkaNenufarIEEE,ZarkaNenuFARinstrument22} observing programmes, we detected anomalous changes in the DM along the LoS to PSR~J1022+1001 during its 2021 solar conjunction. Using white light observations from the LASCO coronagraph and the \textsc{CMEchaser} package, we confirmed that this increase was caused by a coronal mass ejection passing over the LoS. The extreme sensitivity of LOFAR and NenuFAR enabled us to investigate the evolution of the CME plasma and magnetic field in 4 minute time windows.

This monitoring of PSR~J1022+1001 offered a unique opportunity to use a millisecond pulsar as a solar probe. Simultaneous LOFAR and NenuFAR observations strongly confirmed the CME transit across the pulsar's LoS. The high signal to noise ratio of the LOFAR observation facilitated a detailed examination of DM and RM with a 4 minute resolution. The resulting DM time series portrays the evolution of the column electron density in the CME, while the RM time series measures the corresponding changes in the LoS magnetic field. The 4 minute cadence required a correction for the dynamic ionosphere and its RM. The variation of the ionospheric RM was an order of magnitude smaller than the variation in the pulsar RM, which allowed us to derive the average LoS magnetic field $\langle B_{\parallel}\rangle_{\mathrm{PSR}}$ after correcting for interstellar and ionospheric contributions. The pulsar data show an increase of $\langle B_{\parallel}\rangle_{\mathrm{PSR}}$ from $-9~\mathrm{nT}$ to a peak of about $63~\mathrm{nT}$ over the course of the CME crossing.

With the current pulsar timing precision, we demonstrated that a pulsar can provide detailed measurements of a solar phenomenon such as a CME. The resolution obtained on $\langle B_{\parallel}\rangle_{\mathrm{PSR}}$ was set by LOFAR's sensitivity and the length of the sub-integrations. As NenuFAR was still incomplete during the observation, its contribution was mainly limited to the detection of the CME, although its superior DM precision is promising. Future polarization calibration of NenuFAR will significantly improve its ability to measure RM and thus probe the solar environment with a higher precision. 

By comparing these observations with synthetic LoS signatures extracted from a data-constrained 3DCORE reconstruction, we tested the consistency of different flux rope configurations with the pulsar measurements. Among the four configurations considered, only the SWN structure reproduces the observed sign and temporal evolution, while alternative configurations are inconsistent with the data. Although quantitative differences remain between modeled and observed amplitudes, the agreement in sign and temporal evolution shows that millisecond pulsars can serve as remote probes of CME magnetic structure. Future progress will require improved modeling of both magnetic field and density distributions and directly integrating pulsar measurements into CME reconstruction methods.

Overall, this work provides a proof-of-concept demonstration that pulsar observations can be combined with data-constrained CME modeling to test magnetic field configurations remotely. With future observations, improved forward modeling including density structure, and integration of LoS constraints into CME reconstruction, this approach opens a promising path toward improved reconstruction, and eventually forecasting, of CME magnetic structure and their space weather impact.

%%%%%%%%%%%%%%%%%%%%%%%%%%%%%%%%%%%%%%%%%%%%%%%%%%%%%%%%%%%%%%
\begin{acknowledgements}
E.M.Z. gratefully acknowledges support by the Marsden Fund Council grant MFP-UOA2131 from New Zealand Government funding, managed by the Royal Society Te Apārangi. 
    H.T.~R., U.V.~A., C.~M. and E.E.~D. are funded by the European Union (ERC, HELIO4CAST, 101042188). Views and opinions expressed are however those of the author(s) only and do not necessarily reflect those of the European Union or the European Research Council Executive Agency. Neither the European Union nor the granting authority can be held responsible for them. 
    J.~P.~W.~V.~ acknowledges support from the National Science Foundation (NSF) AccelNet award No.~2114721 and by the Deutsche Forschungsgemeinschaft (DFG) through the Heisenberg programme (Project No.~433075039).
    M.D. acknowledges the support by the Croatian Science Foundation under the project IP-2020-02-9893 (ICOHOSS).
    % HORIZON INFRA
    Part of this work was supported by the Project: 101131928 — ACME — HORIZON-INFRA-2023-SERV-01, the Partner Project 6435 of the Science and Technology Center in Ukraine «Peculiarities and interaction of the events in the solar corona and interplanetary medium» and by Ukrainian Program "Scientific and technical (experimental) work in the priority area "Radio physical and optical systems for strengthening the defense capability of the state" for 2025-2026 - "Global monitoring of radio signals of natural and artificial origin of decametre-metre waves in the interests of cosmology and applied problems of defense capability" (state registration number - 0125U000879).
    % NenuFAR sentence
    This paper is partially based on data obtained using the NenuFAR radio-telescope. The development of NenuFAR has been supported by personnel and funding from: Observatoire Radioastronomique de Nan\c{c}ay, CNRS-INSU, Observatoire de Paris-PSL, Université d'Orléans, Observatoire des Sciences de l'Univers en Région Centre, Région Centre-Val de Loire, DIM-ACAV and DIM-ACAV + of Région Ile-de-France, Agence Nationale de la Recherche. 
    % LOFAR sentence
    This paper is based on data obtained with the International LOFAR Telescope (ILT). LOFAR is the Low Frequency Array designed and constructed by ASTRON. It is operated by the European LOFAR ERIC (European Research Infrastructure Consortium) and supported by the national research councils of the participating countries.
    % DE60x
    This article uses data from Unterweilenbach (DE602) LOFAR station funded by the Max-Planck-Institut für Astrophysik, Garching; Tautenburg (DE603) LOFAR station funded by the State of Thuringia, supported by the European Union (EFRE) and BMBF Verbundforschung project D-LOFAR I (grant 05A08ST1); Potsdam (DE604) LOFAR station funded by the Leibniz-Institut für Astrophysik, Potsdam; and Jülich (DE605) LOFAR station supported by BMBF Verbundforschung project D-LOFAR I (grant 05A08LJ1). Observations were carried out in stand-alone GLOW mode, technically operated and supported by the Max-Planck-Institut für Radioastronomie, Forschungszentrum Jülich and Bielefeld University, which also provided computing and storage. Additional financial support came from BMBF D-LOFAR III (grant 05A14PBA), D-LOFAR IV (grant 05A17PBA), and the states of Nordrhein-Westfalen and Hamburg.
    
    % FR606
    LOFAR station FR606 is hosted by the Nan\c{c}ay Radio Observatory and is operated
    by Paris Observatory, associated with the French Centre National de
    la Recherche Scientifique (CNRS) and Université d'Orléans.
    % DSE607
    %% \textbf{Sentence about SE607.}
    % CDN sentence.
    We acknowledge the use of the Nançay Data Center (CDN – Centre de Données de Nançay) facility. The CDN is hosted by the Observatoire Radioastronomique de Nançay (ORN) in partnership with the Observatoire de Paris, the Université d’Orléans, the Observatoire des Sciences de l’Univers d’Orléans (OSUC) and the French Centre National de la Recherche Scientifique (CNRS). The CDN is supported by the Région Centre-Val de Loire (département du Cher). The ORN is operated by the Observatoire de Paris, associated with the CNRS.
    % CCMC sentence
    We acknowledge the Community Coordinated Modeling Center (CCMC) at Goddard Space Flight Center for the use of the Space Weather Database Of Notifications, Knowledge, Information (DONKI), \url{https://kauai.ccmc.gsfc.nasa.gov/DONKI/}.
\end{acknowledgements}

\bibliographystyle{aa}
\bibliography{bibliography}

\appendix
\end{document}